\documentclass[twocolumn]{aastex63}
\usepackage{rotating} 
\usepackage{amsmath,amssymb,bbold,mathrsfs}
\usepackage{subfigure}
\usepackage{multirow}
\usepackage{bm}
\usepackage{color}

\makeatletter
\let\frontmatter@title@above=\relax
\makeatother

\usepackage[nonumberlist,nosuper]{glossaries}
\setacronymstyle{long-short}

\newacronym{rt}{RT}{radiative transfer}
\newacronym{los}{LOS}{lines of sight}
\newacronym{prd}{PRD}{partial frequency redistribution}
\newacronym{hfs}{HFS}{hyperfine structure}
\newacronym{wfa}{WFA}{weak field approximation}
\newacronym{uv}{UV}{ultraviolet}

\begin{document}

\shorttitle{Chromospheric \ion{Mn}{1} resonance triplet}
\title{The circular polarization of the \ion{Mn}{1} resonance lines around 280~nm\\ for exploring chromospheric magnetism}

\author{Tanaus\'u del Pino Alem\'an}
\affil{Instituto de Astrof\'{\i}sica de Canarias, E-38205 La Laguna, Tenerife, Spain}
\affil{Departamento de Astrof\'{\i}sica, Universidad de La Laguna, E-38206 La Laguna, Tenerife, Spain}
\author{Ernest Alsina Ballester}
\affil{Instituto de Astrof\'{\i}sica de Canarias, E-38205 La Laguna, Tenerife, Spain}
\affil{Departamento de Astrof\'{\i}sica, Universidad de La Laguna, E-38206 La Laguna, Tenerife, Spain}
\author{Javier Trujillo Bueno}
\affil{Instituto de Astrof\'{\i}sica de Canarias, E-38205 La Laguna, Tenerife, Spain}
\affil{Departamento de Astrof\'{\i}sica, Universidad de La Laguna, E-38206 La Laguna, Tenerife, Spain}
\affil{Consejo Superior de Investigaciones Cient\'{\i}ficas, Spain}

\begin{abstract}
We study the circular polarization of the \ion{Mn}{1} resonance lines at
279.56, 279.91, and 280.19~nm (hereafter, UV multiplet) by means
of radiative transfer modeling. In 2019, the CLASP2
mission obtained unprecedented spectropolarimetric data in a region of
the solar ultraviolet including the \ion{Mg}{2} h and k resonance lines and
two lines of a subordinate triplet, as well as two \ion{Mn}{1} resonance
lines. The first analysis of such data, in particular those corresponding
to a plage region, allowed the inference of the longitudinal magnetic field
from the photosphere to the upper chromosphere just below the transition
region. This was achieved by applying the weak field approximation to the
circular polarization profiles of the \ion{Mg}{2} and \ion{Mn}{1} lines.
While the applicability of this approximation to the \ion{Mg}{2} lines
was already demonstrated in previous works, this is not the case for the \ion{Mn}{1}
UV multiplet. These lines are observed as absorptions
between the \ion{Mg}{2} h and k lines, a region whose intensity is shaped by
their partial frequency redistribution effects. Moreover,
the only \ion{Mn}{1} stable isotope has nuclear spin $I=5/2$ and thus
hyperfine structure must be, a priori, taken
into account. Here we study the generation and transfer of the intensity and circular
polarization of the \ion{Mn}{1} resonance lines
accounting for these physical ingredients.
We analyze their sensitivity to the magnetic field by means of their
response function, and we demonstrate the applicability of the weak field
approximation to determine the longitudinal component of the magnetic field.
\end{abstract}

\keywords{Solar chromosphere - Radiative transfer - Spectropolarimetry - Magnetic fields}

\section{Introduction}\label{S-intro}

The solar chromosphere extends over 10--20 pressure scale heights between the comparatively
cold and dense photosphere and the million-degree rarefied corona
\citep[see][and references therein]{Carlssonetal2019}. It is within the chromosphere
where, on average, the magnetic pressure finally overcomes the exponentially
decreasing gas pressure and thus the magnetic field starts dominating the dynamics
and structuring of its plasma.
It is thus clear that understanding the chromospheric magnetic field is essential to fully
understanding the physics of this region of the solar atmosphere.

The results of theoretical spectropolarimetric investigations of \gls*{uv} chromospheric
lines \citep[see][and references therein]{Trujilloetal2017} led to the CLASP
\citep[Chromospheric Lyman Alpha SpectroPolarimeter;][]{Kanoetal2012} and CLASP2
\citep[Chromospheric LAyer SpectroPolarimeter;][]{Narukageetal2016} suborbital rocket
experiments. The former successfully measured the intensity and linear polarization
in the hydrogen Lyman-$\alpha$ line in a quiet Sun region \citep{Kanoetal2017}, while
the latter achieved full Stokes measurements in the \gls*{uv} region around the
\ion{Mg}{2} h and k resonance lines (between approximately $279.4$ and $280.6$~nm)
in both a plage \citep{Ishikawaetal2021} and quiet Sun \citep{Rachmeleretal2022}
regions.

The CLASP2 plage observations revealed significant fractional circular polarization
signals of several percent, not only in the \ion{Mg}{2} spectral lines but also in
two of the \ion{Mn}{1} spectral lines in the observed range, namely, two resonance lines
at $279.91$ and $280.19$~nm. The application of the \gls*{wfa} to those profiles
allowed the inference of the longitudinal component of the magnetic field at three
different regions of the solar chromosphere. The inner (outer) lobes of the \ion{Mg}{2}
h \& k lines are sensitive to magnetic fields in the upper (mid) chromosphere
\citep{delPinoetal2020}, while the \ion{Mn}{1} lines are sensitive to the magnetic
field in lower regions of the solar chromosphere. While the height of formation and
applicability of the \gls*{wfa} to the \ion{Mg}{2} lines have been the subject of
several studies \citep{Alsinaetal2016,delPinoetal2016,Afonsoetal2022}, this is not
the case for the \ion{Mn}{1} lines.

In \cite{Ishikawaetal2021} we advanced that, indeed, the \gls*{wfa} can be applied to
these \ion{Mn}{1} lines and that ``A non-LTE radiative transfer investigation
that will be published elsewhere shows that the $V/I$ signals of such \ion{Mn}{1}
lines originate in the lower chromosphere, near the temperature minimum region
in standard solar semi-empirical models''. Here we provide such a demonstration.
The \ion{Mn}{1} resonance lines are observed as absorptions in the spectral
region between the \ion{Mg}{2} h and k lines. That region of the intensity
spectrum is severely affected by \gls*{prd} effects in \ion{Mg}{2}
\citep{Uitenbroek1997}. It is thus necessary to model both multiplets simultaneously,
while accounting for \gls*{prd} effects in \ion{Mg}{2}. Moreover, the only stable
isotope of Mn I has nuclear spin $I=5/2$ and, therefore, \gls*{hfs}
must be, a priori, taken into account.

Some \ion{Mn}{1} lines have been previously studied in the literature,
accounting for the \gls*{hfs}. \cite{Lopezetal2002} and \cite{Asensioetal2007}
studied some of the visible and near-infrared \ion{Mn}{1} lines by assuming
a Milne-Eddington atmosphere.
\cite{BergemannGehren2007} carried out non-LTE intensity calculations,
demonstrating the impact of non-LTE effects on the ionization balance, in
relatively complex \ion{Mn}{1} atomic models. \cite{Bergemannetal2019}
carried out a non-LTE study of the formation of \ion{Mn}{1} lines in late-type
stars, including the development of an atomic model with new and refined
atomic data. However, none of these previous investigations studied the
formation of the \ion{Mn}{1} chromospheric UV lines.

In this work we model the circular polarization profiles of the \ion{Mn}{1}
resonance lines at 279.56, 279.91, and 280.19~nm (hereafter, UV multiplet)
together with the \ion{Mg}{2} h \& k lines,
accounting for \gls*{hfs} in the former and \gls*{prd} effects in the latter.
In Section \ref{S-problem} we briefly describe the physical problem and the
method of solution. In Section \ref{S-Imodel} we describe the atomic models
used in this work and show the height of formation of the \ion{Mn}{1} resonant
multiplet. In Section \ref{S-results} we show the results of our \gls*{rt}
modeling, studying the effect of the \gls*{hfs} and the applicability of the
\gls*{wfa}. Finally, we summarize our conclusions in Section
\ref{S-conclusions}.

\section{Physical problem}\label{S-problem}

We solve the problem of the generation and transfer of polarized radiation to
model the circular polarization in the \ion{Mn}{1} UV multiplet.
\ion{Mn}{1} is a minority species that shows nonlocal thermodynamic
equilibrium effects in its spectral lines radiation. Relatively large model atoms,
in terms of the number of
atomic levels and transitions, are required to correctly compute the population
balance (e.g, \citealt{Bergemannetal2019} for the \ion{Mn}{1} atom;
\citealt{ShchukinaTrujillo2001} for the similar \ion{Fe}{1} atom). We defer
the description of the atomic models to \S\ref{S-Imodel}. There is an additional
complication in the modeling of this multiplet, namely, that the spectral lines are
located in the spectral region between the \ion{Mg}{2} h \& k lines and, while
not blended with the strong emission peaks around the core of these strong
transitions, they are in a region of the spectrum significantly affected
by \gls*{prd} effects in \ion{Mg}{2}. Thus, the \ion{Mn}{1} UV multiplet
appears as absorption lines on the wing pattern of the \ion{Mg}{2} h \& k lines.
Consequently, we need, a priori, to include both atoms in our modeling of the
circular polarization, taking into account \gls*{prd} effects in the
\ion{Mg}{2} h \& k doublet.

\begin{figure*}[htp]
\centering
\includegraphics[width = 0.95\textwidth]{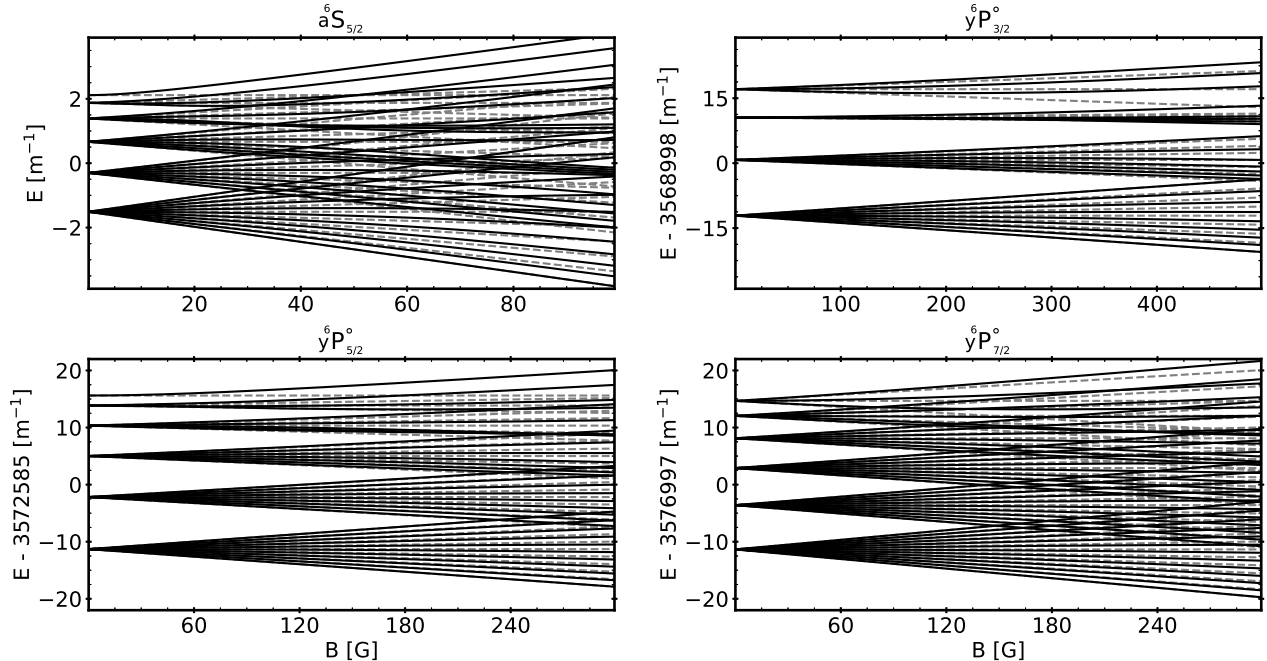}
\caption{The energy of the HFS levels corresponding to each of the atomic levels,
indicated above each panel, of the \ion{Mn}{1} resonance lines, for different
magnetic fields (black solid curves). The gray dashed lines show the same,
but with the linear Zeeman approximation.}
\label{fig::Energygram}
\end{figure*}

Moreover, Mn has only one isotope with nuclear spin $I=5/2$ and, therefore,
\gls*{hfs} must be accounted for. In the presence of a magnetic field, the
\gls*{hfs} levels split into their magnetic components. The \gls*{hfs}
F-levels are relatively close in energy and magnetic field strengths of 10~G
and 80~G are enough to induce repulsion and crossing among the magnetic levels
in the lower (a${}^6$S) and upper (y${}^6$P${}^\circ$) terms, respectively
(see Fig.~\ref{fig::Energygram}). 
Therefore, it is important to consider the general incomplete Paschen-Back effect
regime. Due to the complexity of the problem and the lack of a general \gls*{rt}
code capable of taking into account all the above-mentioned physical ingredients,
we have opted for the following modeling strategy:
\begin{itemize}
\item[1] Compute with the HanleRT code \citep{delPinoetal2016} the population balance
      using relatively extensive atomic models for the \ion{Mn}{1} and \ion{Mn}{2}
      atoms neglecting \gls*{hfs} while including a suitable \ion{Mg}{2} atomic model,
      taking into account \gls*{prd} effects in its resonance transitions.
\item[2] Fix the populations of the two \ion{Mn}{1} terms of the multiplet of interest
      and solve the problem of the
      generation and transfer of polarized radiation for a \ion{Mn}{1} two-term atom
      together with the \ion{Mg}{2} ion applying the HanleRT code.
      We solve the problem with and without accounting for \gls*{hfs}, but always neglecting
      quantum interference between the hyperfine $F$-levels.
\item[3] Taking the resulting radiation field tensors, emissivity and absorptivity
      contributions of the \ion{Mg}{2} h \& k lines, and the population of the
      a${}^6$S \ion{Mn}{1} term, solve the two-term problem with \gls*{hfs}
      and $J-$ and $F-$ state quantum interference with the code developed by
      \cite{Alsinaetal2022}, thus accounting for all relevant physical ingredients,
      including the Paschen-Back effect. Note that for this step we do not solve
      the self-consistent \gls*{rt} problem but instead use the results from the
      second step to obtain a single formal solution along the \gls*{los} of
      interest.
\end{itemize}

\section{Atomic model and height of formation}\label{S-Imodel}

\begin{figure*}[htp]
\centering
\includegraphics[width = 0.82\textwidth]{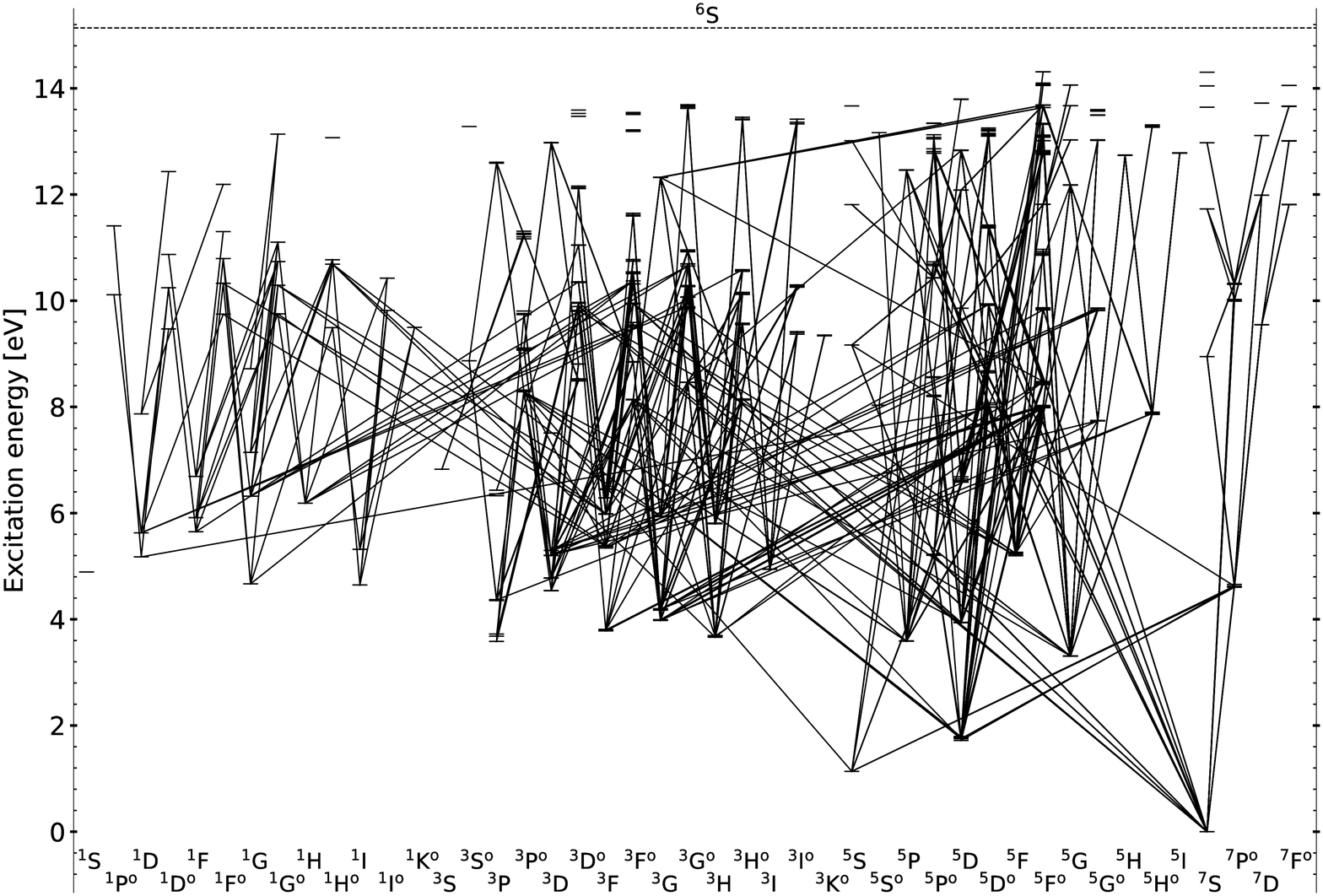} \\
\includegraphics[width = 0.82\textwidth]{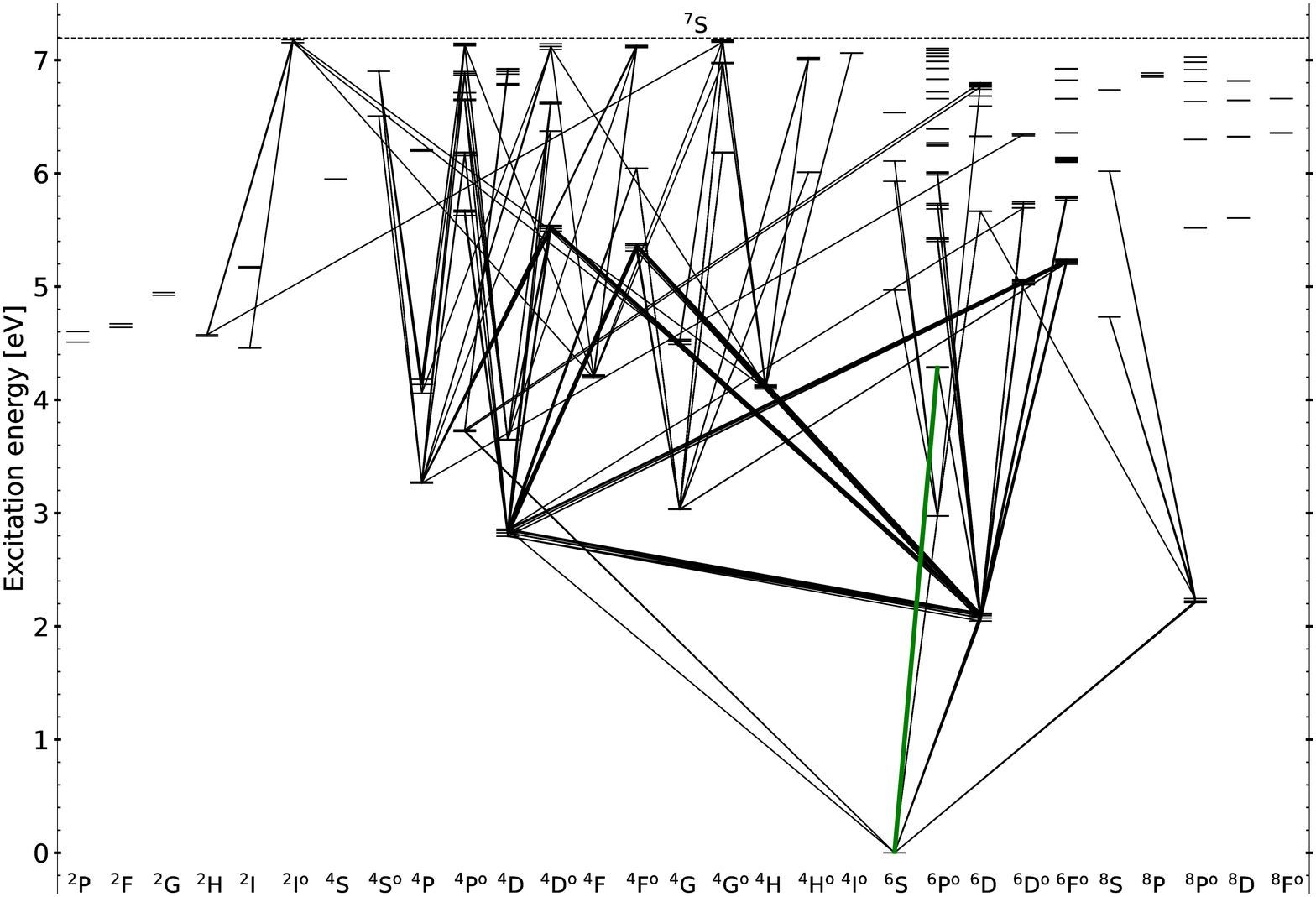}
\caption{Grotrian diagram of the atomic models for the \ion{Mn}{1} (bottom panel)
and the \ion{Mn}{2} (top panel) atoms used to calculate the population balance. The
solid lines show the considered radiative transitions. Every atomic level is connected
to every other level via inelastic collisions with electrons and every \ion{Mn}{1}
level is connected with the ground term of \ion{Mn}{2} via photoionization and
recombination (these transitions are not shown in the diagrams for a better
visualization). The green line shows the transition corresponding to the \ion{Mn}{1}
UV multiplet. The horizontal dashed lines show the ionization energy for
each corresponding atomic species, with the atomic configuration of the ground term of the
next ionization stage.}
\label{fig::MnFullGrotrian}
\end{figure*}

We compute the population balance using an atomic model with
319 \ion{Mn}{1} levels and 318 transitions, 525 \ion{Mn}{2} levels and 891
transitions, and the ground term of \ion{Mn}{3} (see Fig. \ref{fig::MnFullGrotrian}).
These correspond to all levels and transitions available in the NIST database
\citep{NIST}, from where we took the atomic level energies and oscillator
strengths. We accounted for quadratic Stark and collisional broadening in the
spectral lines with the $C_6$ and $C_4$ estimates without enhancements
\citep{BMihalas1978}. We included hydrogenic photoionization cross sections
(e.g., \citealt{BMihalas1978}). The excitation rate due to inelastic
collisions with electrons between levels radiatively connected by an electric
dipole transition was computed following \cite{VanRegemorter1962}, while 
\cite{BelyVanRegemorter1970} was followed for the rest of the level pairs, taking the
parameter $\Omega = 0.1$. Ionizing rates due to inelastic collisions with
electrons were computed following \cite{BAllen1963}. Although these are relatively
rough approximations, they are suitable for the purposes of this paper.

\begin{figure}[htp]
\centering
\includegraphics[width = 0.45\textwidth]{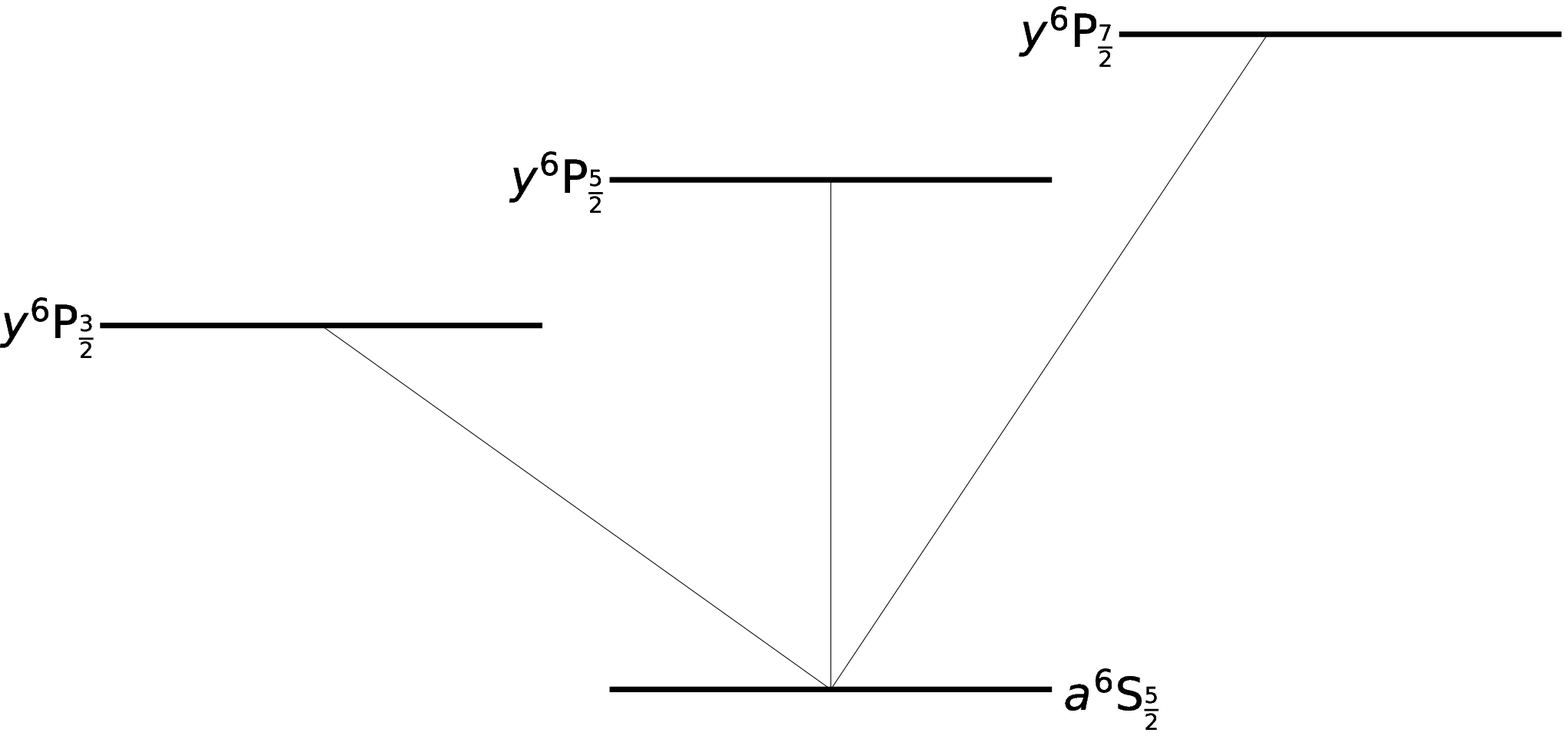} \\
\includegraphics[width = 0.45\textwidth]{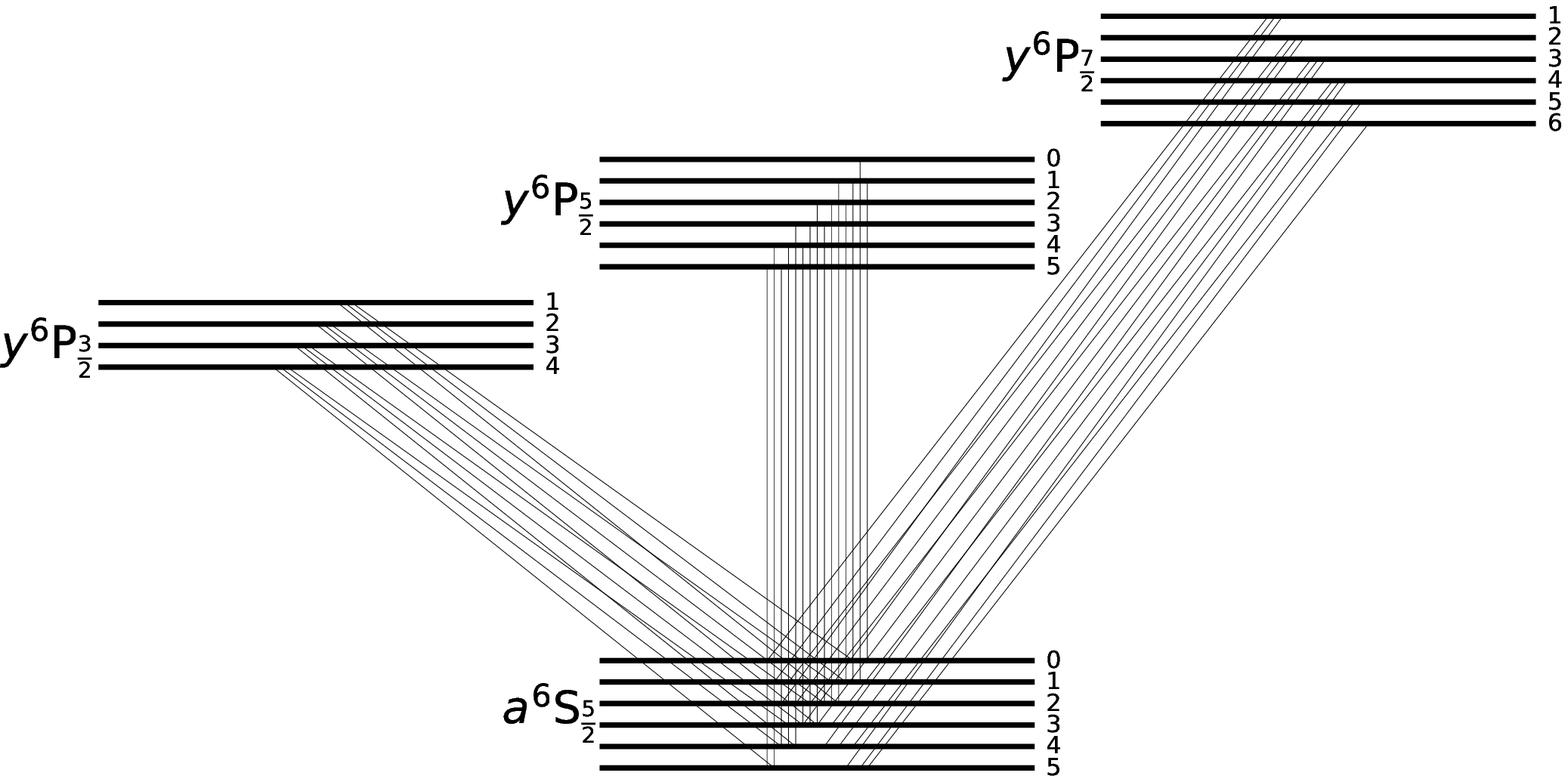}
\caption{Grotrian diagram for the \ion{Mn}{1} UV multiplet neglecting
(top panel) and accounting for (bottom panel) \gls*{hfs}. The four J-levels
radiatively coupled by three transitions become 22 F-levels connected by 42 transitions.}
\label{fig::MnTTGrotrian}
\end{figure}

\begin{figure*}[htp]
\centering
\includegraphics[width=0.97\textwidth]{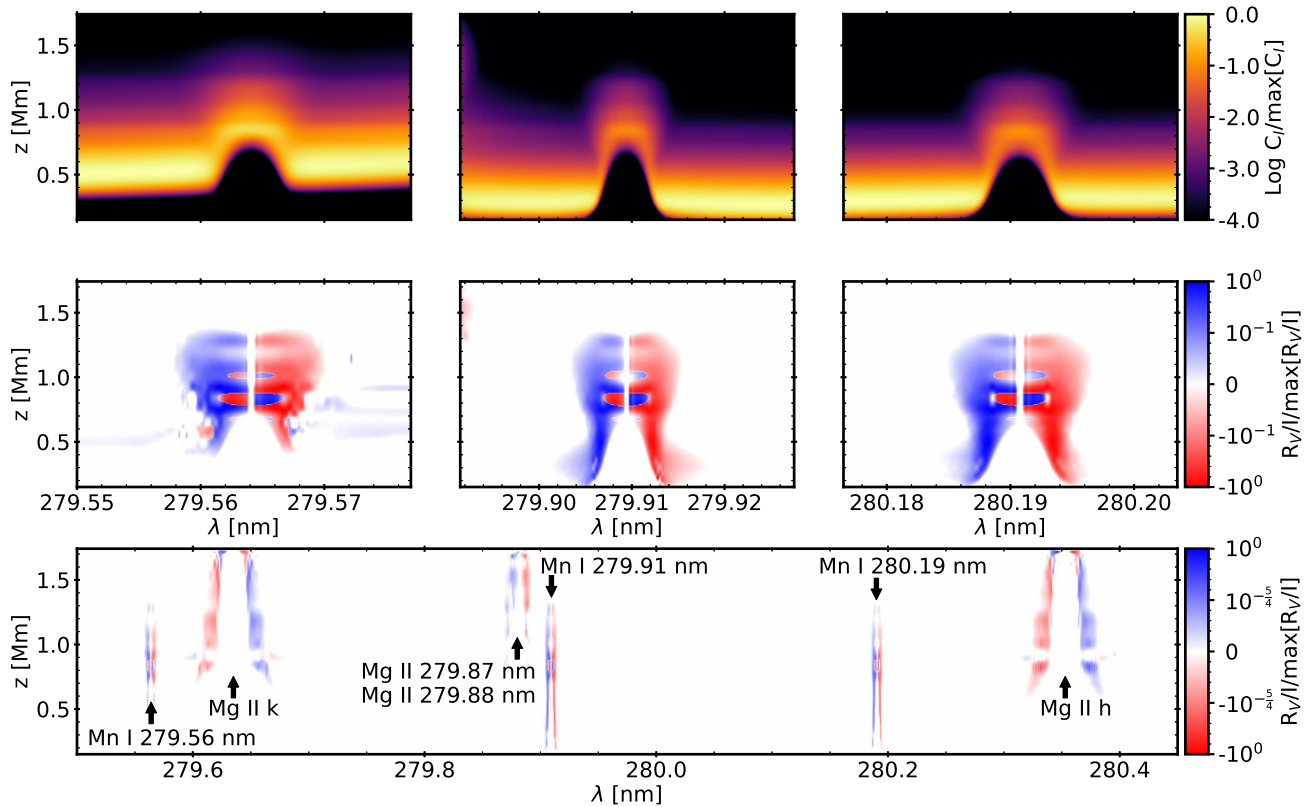}
\caption{Intensity contribution function (top row) and circular polarization
response function (middle row) for the three lines of the \ion{Mn}{1}
UV multiplet (ordered from left to right for increasing wavelength)
in the FAL-P semiempirical model
with a constant vertical 100~G magnetic field, neglecting \gls*{hfs}.
The panel in the bottom row shows
the circular polarization response function for the wavelength range including
the \ion{Mn}{1} UV multiplet and the \ion{Mg}{2} h and k lines and
subordinate lines at 279.88~nm. We show the logarithm of the contribution and response
functions normalized to the maximum for each wavelength range. For the circular
polarization, the red (blue) color indicates that the response to the circular
polarization is negative (positive).}
\label{fig::response-func}
\end{figure*}

The two-term atomic model we use to compute the polarization of the
\ion{Mn}{1} resonance lines includes
the three levels of the upper y${}^6$P term, with total angular
momentum $J=3/2$, $5/2$, and $7/2$, and the single level of the
lower a${}^6$S${}_{5/2}$ term. There are three transitions connecting
each of the levels of the upper term with the lower level (see the left
panel in Fig. \ref{fig::MnTTGrotrian}). We solve the \gls*{rt}
problem for this multiplet in the semiempirical P model of
\citeauthor{Fontenlaetal1993} (\citeyear{Fontenlaetal1993},
hereafter FAL-P) following steps 1 and 2 of the
procedure described in \S\ref{S-problem}. We have chosen the FAL-P
model because it is representative of a plage region, such
as that observed by the CLASP2 mission. In Fig.~\ref{fig::response-func}
we show the intensity contribution function \citep[e.g.,][]{Uitenbroek2006}
for the \ion{Mn}{1} UV multiplet lines, as well as
the circular polarization response
function \citep{Magain1986}, for the same lines and for a spectral
range that also includes the  \ion{Mg}{2} h-k and subordinate triplet lines.
The response function indicates how the circular polarization profile responds
to changes in the magnetic field at each layer in the model atmosphere, and
thus shows where in the model atmosphere is the emergent circular polarization
profile sensitive to the magnetic field. The \ion{Mn}{1} lines at
279.91 and 280.19~nm are thus sensitive to magnetic fields at heights
between about $500$ and $900$~km in the
FAL-P model, which corresponds to the lower chromosphere in that
model. 
The response of the \ion{Mn}{1} line at 279.56~nm is more concentrated
toward the higher part of this range but, unfortunately, this line is
blended with the blue wing of the \ion{Mg}{2} k line.
The faint blue response in the outer wings of the \ion{Mn}{1} line at
279.56~nm corresponds to the \ion{Mg}{2} k line. Note that the
response function for each panel is normalized to the maximum in the
wavelength range of the corresponding panel. In the full wavelength
range (including both the \ion{Mg}{2} resonant doublet and the \ion{Mn}{1}
UV multiplet) the maximum of the response function is in the \ion{Mg}{2} h and
k lines and thus the response function in the bottom panel is normalized
to a value larger than those in the individual panels for each of the
\ion{Mn}{1} lines.
This is the reason why the relatively small positive response in the k line wing
around the \ion{Mn}{1} line at 279.56~nm is seen in the left panel in
the middle row, but not in the bottom panel.
In the calculation of these contribution and response functions
we have neglected \gls*{hfs}, but we do not expect a significant impact
on the height of formation by including its contribution.

When accounting for the \gls*{hfs}, each $J$ level splits into
$(I + J) - |I - J| + 1$ hyperfine levels. The \gls*{hfs} model thus has 22 levels,
with 42 electric dipole transitions (see the right panel in Fig. \ref{fig::MnTTGrotrian}).
The energy for each $F$ level in the model atom with \gls*{hfs} is given by
\cite{Arimondoetal1977} 
\begin{equation}
E_{LJF} = E_{LJ} + A\frac{K}{2} + B\frac{\frac{3}{2}K(K+1) - 2I(I+1)J(J+1)}
                                        {2I(2I-1)2J(2J-1)} ,
\end{equation}
where
\begin{equation}
K = F(F+1) - J(J+1) - I(I+1) ,
\end{equation}
and $A$ and $B$ are the \gls*{hfs} constants given in Table \ref{tab::hfs}.

\begin{table}[htp]
\begin{center}
\caption{Hyperfine structure constants for the atomic levels of the
         \ion{Mn}{1} UV multiplet.}
\label{tab::hfs}
\begin{tabular}{||c | c | r | r | r ||} 
\hline
Term & $J$ & $E_{LJ}$ (cm$^{-1}$) & $A$ (cm$^{-1}$) & $B$ (cm$^{-1}$) \\
\hline\hline
a${}^6$S${}^{(1)}$ & $5/2$ & $0$ & $-2.416$e$-3$ & $-6.348$e$-7$ \\
\hline
\multirow{3}{*}{y${}^6$P${}^{(2)}$}
                    & $3/2$ & $35689.98$ & $-3.24$e$-2$ & $ 1.1$e$-3$ \\\cline{3-5}
                    & $5/2$ & $35725.85$ & $-1.80$e$-2$ & $-2.0$e$-3$ \\\cline{3-5}
                    & $7/2$ & $35769.97$ & $-1.30$e$-2$ & $ 9.0$e$-4$ \\
\hline
\multicolumn{4}{l}{\footnotesize(1) \cite{Davisetal1971}} \\
\multicolumn{4}{l}{\footnotesize(2) \cite{LucGerstenkorn1972}}
\end{tabular}
\end{center}
\end{table}

\vspace{18pt}
\section{Results}\label{S-results}

In this section we show the results of our \gls*{rt} calculations, analyzing
the impact of the \gls*{hfs} on the \ion{Mn}{1} UV multiplet Stokes
profiles, as well as the applicability of the \gls*{wfa} to the circular polarization
profiles to infer the longitudinal component of the magnetic field.


\subsection{Hyperfine structure}\label{SS-HFS}

\begin{figure*}[htp]
\centering
\includegraphics[width = 0.95\textwidth]{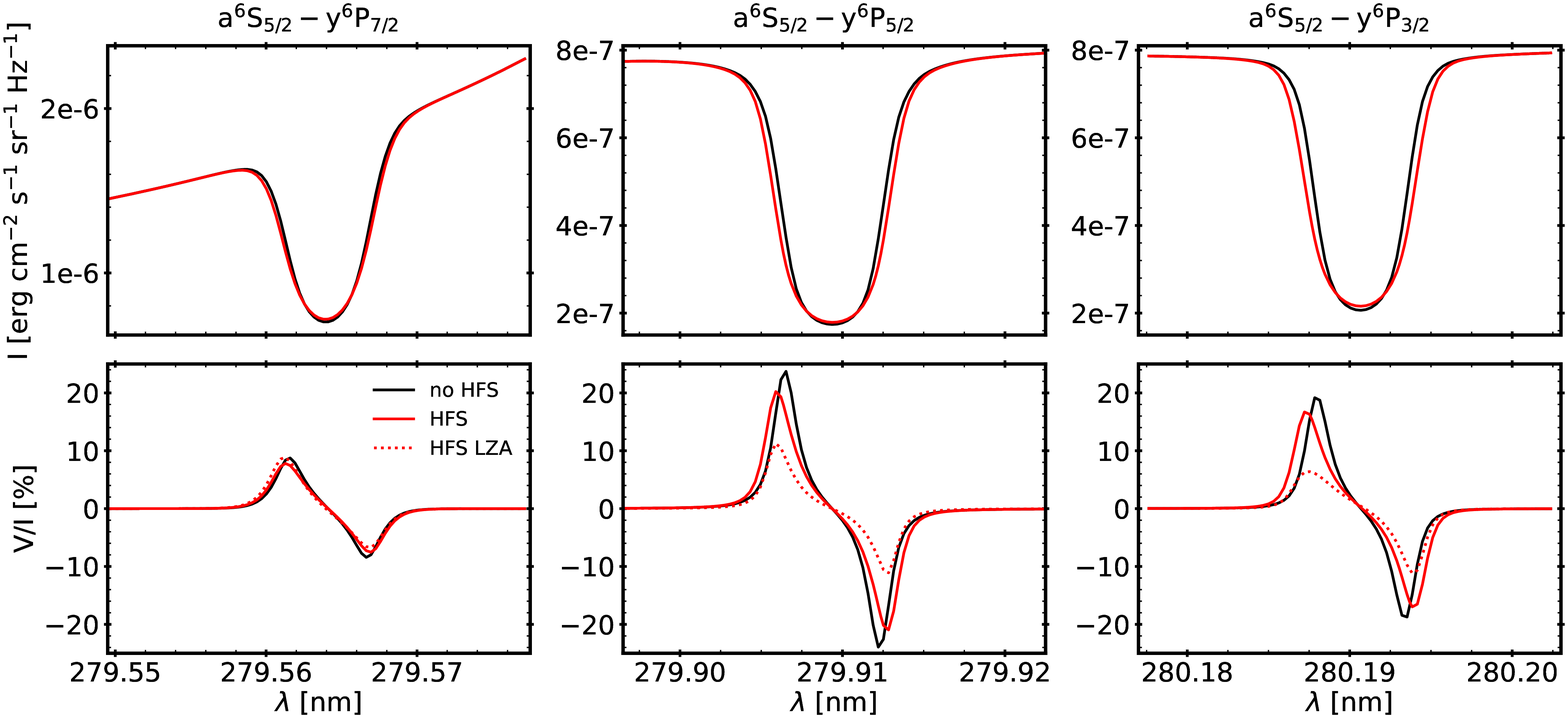}
\caption{Intensity (top panels) and fractional circular polarization $V/I$ (bottom panels)
profiles for a \gls*{los} at the solar disk center, computed in the FAL-P model with a
vertical magnetic field of 500~G following the procedure described in \S\ref{S-problem}.
The different columns show the lines of the \ion{Mn}{1} UV multiplet ordered
by their wavelength from left to right. The black (red) solid curve shows the calculation
neglecting (accounting for) \gls*{hfs}. The red dotted curve shows the circular
polarization profile in the linear Zeeman approximation
(i.e., assuming that the energies of the various magnetic sublevels depend
linearly on the magnetic field strength),
which significantly underestimates
the circular polarization amplitude in two of the \ion{Mn}{1} resonance lines.}
\label{fig::SingleRTE_HFSvsNHFS}
\end{figure*}

We have solved the \gls*{rt} problem for the \ion{Mg}{2} h and k lines,
its subordinate triplet at 279.88~nm, and the \ion{Mn}{1} UV multiplet
following the procedure outlined in \S\ref{S-problem}, both with and without
accounting for \gls*{hfs}, in the FAL-P model permeated by a constant 500~G vertical
magnetic field (see Fig.~\ref{fig::SingleRTE_HFSvsNHFS}). Concerning the intensity profile,
the inclusion of the \gls*{hfs} increases its broadening due to the spread in the energy of the
F-levels. Regarding the circular polarization profile (red curves), not only are the lobes
shifted away from the line center (as expected due to the broader intensity profile), but
their value is decreased with respect to the no-HFS case, as would also be expected if
the circular polarization profile were proportional to the derivative of the intensity.
It is important to emphasize the necessity of accounting for the Paschen-Back
effect to correctly model the circular polarization profile; if the energies of the
magnetic sublevels of the atomic system are assumed to vary linearly with the
magnetic field strength (red dotted curve),
the circular polarization in the
\ion{Mn}{1} lines at 279.91 and 280.19~nm is significantly underestimated.

\subsection{Weak field approximation}\label{SS-WFA}

The main motivation for this investigation is testing the validity of the \gls*{wfa}
to infer the longitudinal component of the magnetic field. To this end, we compare
the circular polarization profile obtained from our \gls*{rt} calculation with
the profile obtained with the \gls*{wfa}, namely \citep[e.g.,][]{BLandiLandolfi2004},

\begin{equation}
V(\lambda) = -4.6686\cdot10^{-13}\,\lambda^2_0\,\bar{g}\,B_{\rm LOS}\,\frac{\partial I}{\partial\lambda} ,
\end{equation}
where $\lambda_0$ is the transition wavelength in \AA, $B_{\rm LOS}$ is the
longitudinal component of the magnetic field in gauss, and $\bar{g}$ is the
effective Land\'e factor of the transition. The $\bar{g}$ values of the
lines of the \ion{Mn}{1} UV multiplet, in wavelength order, are $1.36$, $1.94$, and $1.70$.
Assuming that the contribution
to the Doppler width is purely thermal and assuming a lower boundary of the
temperature of 4000~K, the ratio between Zeeman splitting and Doppler width is,
for the \ion{Mn}{1} line at 279.56~nm (the one with the smallest effective
Land\'e factor), equal or smaller than unity for magnetic fields with
$|B|\lessapprox2$~kG (e.g., this ratio is $\approx0.3$ for $|B|=500$~G). Therefore,
in terms of magnetic field strengths, the \gls*{wfa} approximation is expected
to be applicable to these lines for the typical physical properties of the solar plasma.

\begin{figure*}[htp]
\centering
\includegraphics[width=0.95\textwidth]{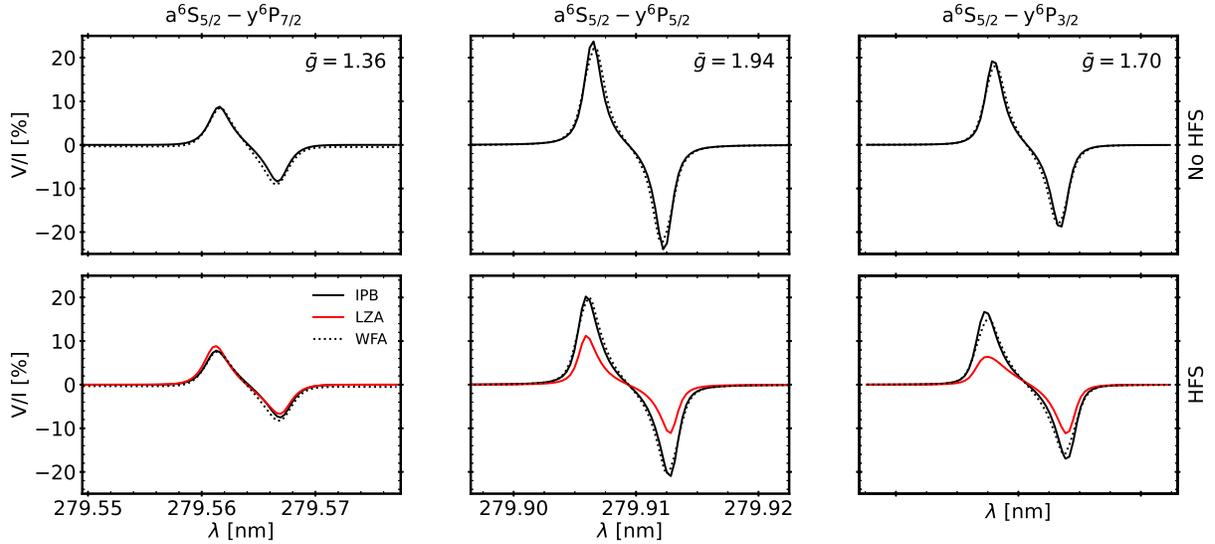}
\caption{Fractional circular polarization $V/I$ profiles for a \gls*{los} at the solar disk
center, computed in the FAL-P model with a vertical magnetic field of 500~G following the
procedure described in \S\ref{S-problem}. The different columns show the lines of the
\ion{Mn}{1} UV multiplet ordered by their wavelength from left to right.
The black solid curves
show the calculation accounting for the incomplete
Paschen-Back effect for fine structure (top panels) and for both fine structure
and \gls*{hfs} (bottom panels).
The red curves in the
bottom panels show the result of the calculation assuming the linear Zeeman approximation
(LZA; i.e., assuming that the energies of the various magnetic sublevels depend linearly
on the magnetic field strength). The black dotted lines show the \gls*{wfa} profile for
the same longitudinal magnetic field and the effective Land\'e factor calculated by
assuming LS coupling and neglecting \gls*{hfs} ($\bar{g}$, indicated in the top panels
for each line).}
\label{fig::SingleRTE_nHFS_Magnetograph}
\end{figure*}

When neglecting \gls*{hfs}, the \gls*{wfa} profiles fit the emergent circular
polarization (see top row of Fig.~\ref{fig::SingleRTE_nHFS_Magnetograph}).
Likewise, when accounting for the \gls*{hfs} (see black curves in the bottom row of
Fig.~\ref{fig::SingleRTE_nHFS_Magnetograph}), the \gls*{wfa} also fits the circular
polarization profile, thus validating its use to infer the longitudinal component
of the magnetic field for weak magnetic fields. However, as advanced in
\S\ref{SS-HFS}, it is important to note
that this fit can only be demonstrated if the \gls*{rt} problem is solved accounting
for all the relevant physical ingredients, namely \gls*{hfs} in the incomplete
Paschen-Back regime.
To illustrate the latter, we have also included the \ion{Mn}{1} multiplet circular
polarization profiles calculated in the linear Zeeman regime (see red curves in the
bottom row of Fig.~\ref{fig::SingleRTE_nHFS_Magnetograph}). In the third step of the modeling
(see Sect.~\ref{S-problem}) this regime was attained by neglecting
the off-diagonal elements of the magnetic Hamiltonian, so that the energies of the
magnetic sublevels vary linearly with the magnetic field strength.
Under such an approximation the circular polarization profile is underestimated and
one could reach the wrong conclusion that the \gls*{wfa} profile does not fit the
circular polarization. It is also important to emphasize
that the \gls*{wfa} profiles in the bottom row of Fig.~\ref{fig::SingleRTE_nHFS_Magnetograph}
(with \gls*{hfs}) have been calculated using the effective Land\'e factors of the transition
neglecting the \gls*{hfs}. This is an example of the principle of spectroscopic stability:
``if two different descriptions are used to characterize a quantum system -- a detailed
description which takes an inner quantum number into account and a simplified description
which disregards it -- the predicted results must be the same in all physical experiments where
the structure described by the inner quantum number is unimportant'' \citep{BLandiLandolfi2004}.
The differences in energy between the \gls*{hfs} F-levels of each J-levels are thus small
enough to satisfy this principle.

\section{Conclusions}\label{S-conclusions}

We have studied the formation of the circular polarization profiles of the
\ion{Mn}{1} UV multiplet taking into account its \gls*{hfs}
as well as \gls*{prd} effects in the \ion{Mg}{2} h and k lines that shape
the intensity ``continuum'' on top of where the \ion{Mn}{1} absorptions are
observed. From the response function to perturbations to the longitudinal
component of the magnetic field we have shown that the circular polarization
of the \ion{Mn}{1} resonant triplet is sensitive to the magnetic field in
a more or less extensive region around $900$~km in the FAL-P model atmosphere,
which corresponds to its lower chromosphere. This justifies the decision to
assign the magnetic field inferred from the application of the \gls*{wfa}
to the \ion{Mn}{1} circular polarization profiles to the lower chromosphere
in \cite{Ishikawaetal2021}.

While the \gls*{hfs} energy splitting slightly broadens the \ion{Mn}{1} triplet
intensity profiles, its impact on the circular polarization profiles is
more evident, shifting its lobes further from the line core and slightly
reducing their amplitude. We have demonstrated that the linear Zeeman
approximation is not suitable to model the \ion{Mn}{1} circular polarization
profiles with \gls*{hfs}, as it severely underestimates the polarization
amplitude.

Finally, the circular polarization profiles obtained by applying the
\gls*{wfa} with the known homogeneous magnetic field and the effective
Land\'e factors corresponding to the fine structure transitions
(i.e., as if neglecting \gls*{hfs}) fit the circular polarization profiles
resulting from the calculations accounting for the incomplete Paschen-Back
effect. This demonstrates that the \gls*{wfa} can be used to infer the
longitudinal magnetic field component without accounting for \gls*{hfs}
because these lines fulfill the principle of spectroscopic stability; i.e.,
because the \gls*{hfs} energy split is sufficiently small.

We want to emphasize that because we are imposing a homogeneous magnetic
field in a static model atmosphere, physical conditions under which the
\gls*{wfa} should work, what we have demonstrated is that the \gls*{wfa}
is applicable to these lines despite the \gls*{hfs} and their potential
interaction with the \ion{Mg}{2} h and k wings. Of course, if the magnetic
field gradient is too steep across the region of formation, we would run
into the usual limitations of the \gls*{wfa} as an inference tool, namely,
we would infer a particular value for the longitudinal magnetic field, but
we would be unable to pinpoint where along the \gls*{los} is such a field
localized. However, we can expect that such a magnetic field is at least
representative of the magnetic field somewhere across the region of formation
of the line, as in \cite{Ishikawaetal2021}.

\acknowledgements
We acknowledge the funding received from the European Research Council (ERC)
under the European Union's Horizon 2020 research and innovation program (ERC
Advanced grant agreement No. 742265).

\bibliographystyle{apj}
\bibliography{apj-jour,biblio}


\end{document}